\documentclass[journal]{IEEEtran}
\usepackage{amsmath,graphicx}
\usepackage{psfrag}
\usepackage[linesnumbered,lined,boxed,commentsnumbered]{algorithm2e}
\IncMargin{1em}
\usepackage{lineno}
\usepackage{url}
\usepackage{xcolor}
\modulolinenumbers[5]

\usepackage[hidelinks]{hyperref}

  \usepackage[absolute]{textpos}
\newcommand{\copyrightstatement}{
    \begin{textblock}{15}(0.5,0.3)    
         \noindent
         \centering
         \textblockcolour{white}
         \footnotesize
         \copyright 2021 IEEE. Personal use of this material is permitted. Permission from IEEE must be obtained for all other uses, in any current or future media, including reprinting/republishing this material for advertising or promotional purposes, creating new collective works, for resale or redistribution to servers or lists, or reuse of any copyrighted component of this work in other works
    \end{textblock}
}

\begin{document}
\title{Component-wise Power Estimation of Electrical Devices Using Thermal Imaging }
\author{Christian~Herglotz, Simon~Grosche, Akarsh~Bharadwaj, and Andr\'e~Kaup
\thanks{Manuscript received March 22, 2021; revised August 16, 2021, accepted September 27, 2021.}
\thanks{C. Herglotz, S. Grosche, A. Bharadwaj, and A. Kaup are with the Chair of Multimedia Communications and Signal Processing,  Friedrich-Alexander University Erlangen-N\"urnberg (FAU), Cauerstr. 7, 91058 Erlangen, Germany. Email: \{christian.herglotz, simon.grosche, akarsh.bharadwaj, andre.kaup\}@ fau.de}}

\copyrightstatement

\maketitle

\begin{abstract}
This paper presents a novel method to estimate the power consumption of distinct active components on an electronic carrier board by using thermal imaging. The components and the board can be made of heterogeneous material such as plastic, coated microchips, and metal bonds or wires, where a special coating for high emissivity is not required. The thermal images are recorded when the components on the board are dissipating power. In order to enable reliable estimates, a segmentation of the thermal image must be available that can be obtained by manual labeling, object detection methods, or exploiting layout information. Evaluations show that with low-resolution consumer infrared cameras and dissipated powers larger than $300$\,mW, mean estimation errors of $10\%$ can be achieved. 
\end{abstract}


\IEEEpeerreviewmaketitle

\section{Introduction}
\label{sec:intro}

\IEEEPARstart{I}{n} recent years, the world-wide use of small consumer electronic devices such as smartphones, tablet PCs, processing boards, or field-programmable gate arrays (FPGAs) has increased dramatically. 
These devices are used in versatile applications such as monitoring, surveillance, detection, smart home, and IoT. In many applications, the devices are connected to the internet and perform multiple tasks in parallel. As many of these devices are battery-driven, the energy consumption of such a device attracts more and more attention. {The goal is to extend operating times and to keep the overall energy consumption at a reasonable overall level. }
The energy consumption of such devices was assessed and optimized in many studies, for example for smartphone apps in general \cite{Carroll10,Carroll13,Li17bigLITTLE}, in video coding  \cite{Herglotz18,Herglotz19,Mallikarachchi20,Singhadia20}, video streaming \cite{Herglotz19b,Herglotz20,Diaz20}, wearable consumer devices \cite{Hong20}, or smart home applications \cite{Gray20,Buddhahai20}.   

A common challenge in the design of electronic circuit boards and algorithms is the optimization of the processing complexity and the power demand of the target device, where the power can be consumed by multiple distinct hardware components on the board. During the development of new boards and algorithms, an important task is to test and analyze both for their power demand in order to allow power optimizations. Unfortunately, state-of-the-art tools are restricted to either a global or a local analysis. 

In the case of a global analysis, measuring the energy and power demand of the complete device-under-test (DUT) with a power meter can be performed as shown in \cite{Herglotz18,Li12}. As a drawback, this kind of analysis does not provide information on the specific power sinks on the DUT. For example, in a surveillance scenario in which the camera uploads video data in real-time to the internet, the overall power would be distributed among multiple sinks such as the camera sensor, the video encoder chip, the CPU for flow control, and the network chip. 

In the case of a local analysis, each component of the DUT is analyzed separately. On the algorithmic level, this can be achieved performing a complexity or time analysis of the implementation offline \cite{Elabidine14,Bossen12}. As a drawback, each software component must be analyzed separately and hardware components such as the camera or the network adapter are not considered at all. For power or energy measurements on the hardware level, each component can be measured separately probing the voltage supply of components as performed in \cite{Carroll10, Carroll13}. As a drawback, such a setup is expensive and time-intensive to build and it is practically impossible to cover all power sinks on a modern, heterogeneous electronic circuit board, which includes a large number of distinct hardware components such as SoCs, sensors, actors, shunts, capacitors, etc. Furthermore, the system itself is changed by the incorporation of shunts. The method proposed in this paper targets a local analysis of power consumption with a much simpler and cheaper setup {using thermal imaging}, where the measured system is not changed by additional measurement circuitry.

Next to power sinks such as resistors or processor chips, the proposed method also allows the measurement of the power losses in switch mode power supplies. In practice, it is not feasible to determine the power loss in active as well as magnetic components inside the power supply by measuring voltage and current waveforms. For active components during the on-stage, the voltage is near zero while the current is high. In the off-stage, the blocking voltage reaches very high values while the current is close to zero. Only at the switching instances, in very short periods of time, voltage and current are both present leading to a direct power dissipation both in active and reactive power, as parasitic capacitances are charged or discharged. For magnetic components, the same holds true as the phase shift between current and voltage is close to $\frac{\pi}{2}$ and thus, only a small fraction of apparent power is dissipated. With the proposed measurement setup, power estimates of such components will be enabled.

In order to perform such measurements, we are using a consumer thermal camera that can be attached to smartphones. 
In recent years, much research on thermal cameras has been performed with the goal of developing new applications, which lead to interesting solutions and use cases. 
Examples are the detection of faults in photovoltaic modules \cite{Lee18,Henry20} and their shelters \cite{Tzeng17}, the assessment of the status of grapevine water in biosystems engineering \cite{Petrie19}, and allowing {contact-free} touch interaction on screens \cite{Lee17thermal}. With this paper, we show that consumer thermal cameras can also be used for power estimations on consumer electronic devices.

A similar approach of power estimation using thermal images targeting a single integrated chip (IC) was proposed by Hamann et al. \cite{Hamann07}. They used a method called spatially-resolved imaging of microprocessor power (SIMP), which records thermal images of a microprocessor chip and maps the temperature values to power dissipation values in a steady state. {As a basic principle, this approach exploits the duality between heat conduction and electricity, which states that temperature differences and power flow can be handled equivalently to voltage and current, respectively \cite{Krum00}. } 

Later, Qi et al. \cite{Qi10} applied SIMP to a silicon chip with active as well as inactive regions and incorporated a model to consider transient effects. Then, Paek et al. \cite{Paek13} introduced a probabilistic model into SIMP and showed that the power dissipation of a field-programmable gate array (FPGA) can be modeled accurately. {Finally, Sadiqbatcha et al. introduced a thermal-imaging based method to locate heat sources on a modern CPU \cite{Sadiqbatcha21}}.  

In contrast to the aforementioned work, we not only consider a single chip but a heterogeneous board that can carry multiple distinct components{, which, to the best of our knowledge, was not done before}. Such boards are often major elements of consumer electronic devices. Examples are printed circuit boards (PCB){, smartphones,}  or other electronic circuitry, which, e.g., implement the functionality of a general purpose PC. { A proof-of-concept study on power estimation using thermal imaging found that the overall power consumption of a smartphone is correlated with the heatmap from a thermal camera \cite{Huber19,Huber20}. However, a algorithmic solution for the estimation of the actual power was not provided. } 

\begin{figure}[tb]
\psfrag{M}[c][b]{Component map $\mathcal{M}$}
\psfrag{R}[l][l]{IR-camera}
\psfrag{C}[c][c]{Carrier board}
\psfrag{P}[c][c]{Power estimation}
\psfrag{A}[c][c]{$0.8$ W}
\psfrag{B}[c][c]{$0.7$ W}
  \centering
	\includegraphics[width=.48\textwidth]{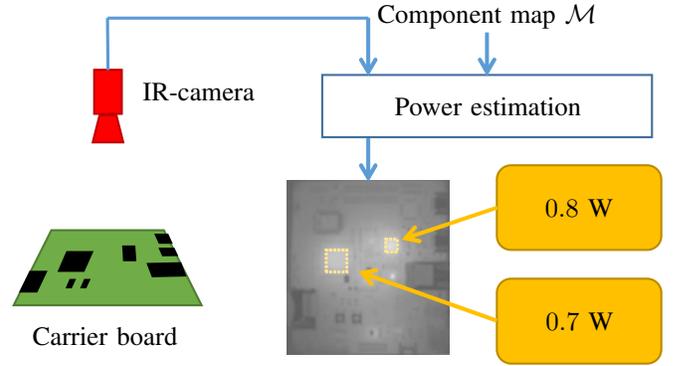}
	\caption{Measurement setup with exemplarily estimated power output. The IR-camera sends a temperature map to the power estimation algorithm, which assigns estimated powers to the distinct components (component-wise power estimation). The distinct components are described by a 2D-map $\mathcal{M}$ at the input of the algorithm. }
\label{fig:setup}
\end{figure}

{In our setup, we consider multiple components on an evaluation board, namely } system-on-chips (SoCs), power management integrated circuits (PMICs), or Wi-Fi chips, {which we monitor } with a single infrared (IR)-camera. The general setup of the method is visualized in Fig.~\ref{fig:setup}. 

To approach this goal, we exploit a simplified test setup in a controlled environment. We measure the power dissipation of resistors on a carrier board, where the true power consumption of each component is known a-priori such that there is no need for sophisticated power measurement circuitry. Still, different materials (resistors, board plastic, metal wires) are part of the setup, which is similar to PCBs consisting of a carrier board, different ICs, and various metal connections. Moreover,  we only consider the steady-state power consumption such that transient aspects are neglected. 
{
 The proposed energy estimation setup provides the following advantages: 
\begin{itemize}
\item cheap and simple setup using a single infrared camera, 
\item no complicated measurement circuitry needed, 
\item no impact on the system due to shunts. 
\end{itemize}
As such, this paper presents the following novel contributions, which have not been proposed before: 
\begin{itemize}
\item power estimation using thermal imaging for heterogeneous carrier boards, 
\item exploitation of a pixel map indicating the type of the recorded component on the carrier board,
\item simplification of an existing method targeting single chips \cite{Hamann07}, 
\item extension of the model by variable emissivity, heat flux, and heat radiation, 
\item estimation of power values without the need of coating to increase emissivity. 
\end{itemize}
}
\noindent As an output of the proposed method, a power estimate for each hardware component on the carrier board is obtained.

The paper is organized as follows. First, Section~\ref{sec:theory} develops our proposed algorithm and discusses the theoretical background. Then, Section~\ref{sec:setup} presents our measurement setup in detail. Afterwards, Section~\ref{sec:eval} evaluates the proposed power estimation method and 
Section~\ref{sec:concl} concludes this paper. 

\section{Temperature-to-Power Conversion}
\label{sec:theory}
{ 
The algorithm proposed in this work is based on the aforementioned SIMP \cite{Hamann07}, which models heat transfer on a discretized two-dimensional surface. In this model, the relation between the power dissipation at a certain pixel position and the temperature is given as}
\begin{equation}
\begin{pmatrix}
a_{11,11} & a_{11,12} & ... & a_{11,HW} \\
a_{12,11} & a_{12,12} & ... & a_{12,HW} \\
\vdots & \vdots & \tiny{\ddots} & \vdots \\
a_{HW,11} & a_{HW,12} & ... & a_{HW,HW} \\
\end{pmatrix}
\begin{pmatrix}
p_{11}\\
p_{12}\\
\vdots\\
p_{HW}
\end{pmatrix}
=
\begin{pmatrix}
T_{11}\\
T_{12}\\
\vdots\\
T_{HW}
\end{pmatrix},
\label{eq:SIMP}
\end{equation}
where $p_{mn}$ is the electrical power of an active component { at position $(m,n)$ on the surface of the microchip, which we call `cell' in the following. In case of thermal imaging, $(m,n)$ are the pixel coordinates. } $T_{mn}$ is the corresponding temperature, $H$ and $W$ are the {vertical and the horizontal number of cells}, respectively. The matrix entries $a_{mn,ij}$ are interpreted as the influence of the power dissipation of cell $(i,j)$ on the temperature of cell $(m,n)$. 

{ For this model, it is assumed that there is an IR-transparent cooling system placed on top of the microchip-under-test \cite{Hamann07}, which provides a predefined and constant cooling rate. The entries $a_{mn,ij}$ are determined by calibration, where a laser beam is used to heat the surface of the microchip at all cell positions with subsequent IR imaging in the steady state. The method that we propose in this paper eliminates the need for a cumbersome calibration step.  }

SIMP was designed for a perfectly flat surface such that the area corresponding to each pixel is constant. However, the surface of electronic boards usually incorporates 3D structures. In our approach, we assume that the height of such 3D structures on electronic boards is generally much smaller than the width and the length of the board. As such, the assumption of a flat DUT is still valid, which will be validated by the evaluation in Section~\ref{sec:eval}.

{Furthermore, we extend SIMP by taking into account that power flow can also occur due to dissipation  to the ambient air \cite{Penoncello19} and due to radiation \cite{Rogalski19}. Based on these basic considerations, we develop our proposed algorithm by first simplifying the SIMP model for the 2D power flow }in Subsection~\ref{secsec:prune}. In SIMP, it is assumed that the input temperatures $T_{mn}$ represent the true temperature of the microprocessor at the pixel position. In our application, temperature readings can be erroneous because of reflective surfaces {with variable emissivities}. This problem is discussed in Subsection~\ref{secsec:emiss}. Afterwards, we present a pixel-wise model for power estimation in Subsection~\ref{secsec:pixModel} that makes use of the ambient room temperature, which is discussed in Subsection~\ref{secsec:ambTemp}.  Finally, the complete component-wise estimation method is presented in Subsection~\ref{secsec:compPower}. 

\subsection{Parameter Pruning}
\label{secsec:prune}

\begin{figure}[t]
\psfrag{T}[c][c]{$C_\mathrm{T}$}
\psfrag{B}[c][c]{$C_\mathrm{B}$}
\psfrag{L}[c][c]{$C_\mathrm{L}$}
\psfrag{R}[c][c]{$C_\mathrm{R}$}
\psfrag{P}[l][l]{\hspace{-.2cm} $p_\mathrm{c},p_\mathrm{rad}$}
  \centering
	\includegraphics[width=.48\textwidth]{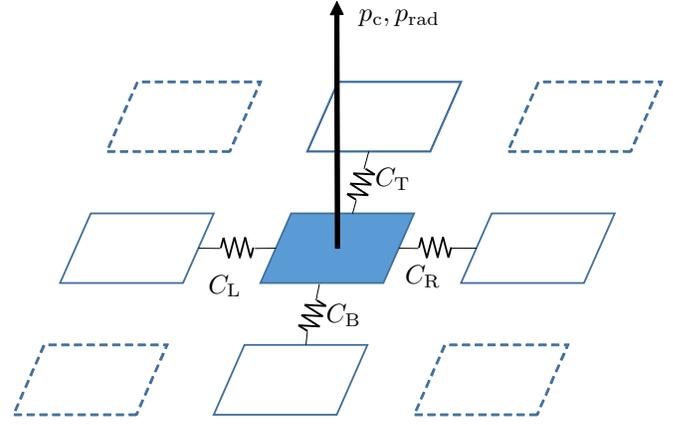}
	\caption{Thermal conductance $C_\mathrm{T}$, $C_\mathrm{B}$, $C_\mathrm{L}$, and $C_\mathrm{R}$ between {cells} (blue). The black arrow indicates the convective power $p_\mathrm{c}$ and the radiative power $p_\mathrm{rad}$, which are dissipated through the air and through thermal radiation, respectively.  }
\label{fig:thermResis}
\end{figure}

{We reduce the number of parameters from \eqref{eq:SIMP} following the aforementioned duality between heat transfer and electricity \cite{Krum00}. To this end, we interpret the parameters $a_{mn,ij}$ as thermal resistances between two cells} as done in \cite{Paek13}. Figure~\ref{fig:thermResis} illustrates this concept for an arbitrary {cell}, where we use the inverse of the thermal resistance, the thermal conductance $C$, to avoid confusion with the electrical resistance  $R$ used in Section~\ref{sec:setup}. 
In this representation, each parameter $a_{mn,ij}$ can be calculated aggregating all conductance paths between the two {cells} at positions $(m,n)$ and $(i,j)$. As such, we claim that it is sufficient to define the thermal conductance between adjacent {cells}. Furthermore, we only consider the four nearest neighbors 
 and neglect diagonally neighboring {cells}. Hence, we obtain the four thermal conductance values $C_\mathrm{T}$, $C_\mathrm{B}$, $C_\mathrm{L}$, and $C_\mathrm{R}$ for transitions to the top, the bottom, the left, and the right {cell}, respectively. With this approach, only $2\cdot (H-1)\cdot(W-1)$ parameters for the vertical and the horizontal conductance values remain. 

In the next step, we consider that the thermal conductance depends on the material linking the two {cells}. For example, we can expect that the conductive tracks have a smaller thermal resistance (i.e. a higher thermal conductivity) than the carrier board material. However, we assume the thermal conductance of a specific material to be constant. 

We further assume that a {cell} cannot cover multiple components on the board, which means that the material corresponding to one {cell} is uniquely defined. Consequently, all conductance values connecting {cells of }the same material are assumed to be equal. 

In our setup, we consider $\mu =3$ different materials (board, wires, resistors) and obtain
$\sum_{i=1}^{\mu=3} i=6$ different conductance values, which are  $C_{11}$, $C_{22}$, and $C_{33}$ when the neighboring {cells} 
 represent the same material; and $C_{12}=C_{21}$, $C_{13}=C_{31}$, as well as $C_{23}=C_{32}$ when the neighboring {cells} represent different materials. These values are collected in the symmetric matrix
 \begin{equation}
 \mathcal{C} = \begin{pmatrix}
C_{11} & C_{12} & ... & C_{1\mu}\\
\vdots & \vdots & \ddots & \vdots \\
C_{\mu1} & C_{\mu2} & ... & C_{\mu\mu}\\
\end{pmatrix}, 
\label{eq:symmetricConductanceMatrix}
 \end{equation}
which can be defined for an arbitrary number of materials $\mu$. The matrix is used later to solve the temperature-to-power conversion problem.

\subsection{Emissivity}
\label{secsec:emiss}
In order to allow temperature readings of bodies, thermal image detectors exploit the Stefan-Boltzmann law that reads 
\begin{equation}
E = \sigma \cdot T^4, 
\label{eq:boltz}
\end{equation}
where $T$ is the temperature in Kelvin of the body's surface, $\sigma\approx5.67 \cdot 10^{-8}\frac{\mathrm{W}}{\mathrm{m^2K^4}}$ is the Stefan-Boltzmann constant, and $E$ is the radiant exitance of the body with units $\left[\frac{\mathrm{W}}{\mathrm{m^2}}\right]$  \cite{Rogalski19}. The exitance $E$ is sensed on the thermal sensor and used to calculate the temperature $T$ of the body { by inverting \eqref{eq:boltz}}. 

Unfortunately, \eqref{eq:boltz} only holds for black bodies. For practical bodies, the observable radiant exitance is given by \cite{Rogalski19}
\begin{equation}
E = \varepsilon\cdot\sigma \cdot T^4 + (1-\varepsilon)\cdot E_\mathrm{in}, 
\label{eq:emiss}
\end{equation}
where $\varepsilon\in[0,1]$ is the emissivity of the body, which describes the fraction of emitted exitance $E$ caused by the temperature of the body $T$. In contrast, $1-\varepsilon$ is the reflectance of the body, which describes the fraction of emitted exitance $E$ caused by reflections of the incident radiation $E_\mathrm{in}$. In our setup, all materials are grey bodies with $\varepsilon<1$, such that the { incorrect temperature measurements need to be compensated}. 

Regarding a carrier board and its components, we assume that it consists of three main materials. The first material constitutes active components based on silicon, which are usually coated with black plastic (e.g., ICs). The second material constitutes the carrier plastic, which is often green or dark green colored. Both materials have a high emissivity close to one. In contrast, the metal wires being the third material usually have a small emissivity { due to their high reflectance }(e.g., $\varepsilon\approx 0.2$ for polished iron \cite{Rogalski19}). As a consequence, measured temperature values for {cells} corresponding to the metal wires also depend on the reflected radiation $E_\mathrm{in}$ as shown in \eqref{eq:emiss}.  { For each material, we define a single constant emissivity value, which is listed in the 3D vector $\mathcal{E}$.}

{ To consider emissivity in our proposed algorithm, we propose two solutions to find the most practical approach. For the first solution, we assume that the emissivity of the metal wires and the incident radiation are constant ($\varepsilon=\mathrm{const}$, $E_\mathrm{in}=\mathrm{const}$). For the latter assumption, we have to make sure that all objects in the visible vicinity of the carrier board have the same temperature (e.g., a constant ambient temperature). This is assured by using a covering box around the board and the thermal camera as will be explained in Section~\ref{sec:setup}. With these assumptions, using \eqref{eq:boltz} we can rewrite \eqref{eq:emiss} to obtain the true temperature of the wires as
\begin{equation}
T_\mathrm{wire} = \sqrt[4]{\frac{\ T_\mathrm{in}^4-(1-\varepsilon_\mathrm{wire})\cdot  T_\mathrm{box}^4 }{\varepsilon_\mathrm{wire}}}, 
\label{eq:temp_modify}
\end{equation}
where $T_\mathrm{in}$ is the temperature that is observed by the camera, $\varepsilon_\mathrm{wire}$ is the constant emissivity of the wires, and $T_\mathrm{box}$ is the temperature of the surrounding box, whose radiation is $E_\mathrm{box}=E_\mathrm{in}$. It is worth mentioning that for emissivities close to zero, small deviations in the emissivity $\varepsilon_\mathrm{wire}$ can lead to strong variations in the calculated temperature $T_\mathrm{wire}$. 

For the second solution, }we assume that the temperature distribution in the DUT is homogeneous and follows a spatial low-pass behavior. In our case, we can use the measured temperatures in the surrounding area of the wires 
to estimate the true temperature of the wires. Using the map of components $\mathcal{M}$ that provides information about the pixels representing wires, we remove the corresponding temperature information from the observed temperature map and interpolate the temperature of the wires using bicubic interpolation.

\subsection{Pixel-wise Power Model}
\label{secsec:pixModel}
{ In the steady state}, the sum of incoming and outgoing power { of each cell} (the power flow) must be zero as
\begin{equation}
p_\mathrm{el}  + p_\mathrm{T} + p_\mathrm{B} + p_\mathrm{L} + p_\mathrm{R} + p_\mathrm{c}+ p_\mathrm{rad} = 0. 
\label{eq:pixelModel}
\end{equation}
In this equation, $p_\mathrm{el}$ is the electrical power, i.e. the electrically dissipated power, which shall be estimated. The powers $p_\mathrm{T}$, $p_\mathrm{B}$, $p_\mathrm{L}$, as well as $p_\mathrm{R}$ represent the power flow through the thermal conductances in Fig.~\ref{fig:thermResis}.  $p_\mathrm{c}$ is the convective power, which is the cooling power dissipated to the ambient room. 

The radiative power $p_\mathrm{rad}$ is the difference between the emitted radiance \eqref{eq:emiss} and the absorbed incoming radiance given by 
\begin{equation}
p_\mathrm{rad} = A\cdot\left( \varepsilon \cdot \sigma\cdot T_0^4 -\varepsilon \cdot E_\mathrm{in}\right),
\label{eq:radPower}
\end{equation}
where $A$ is the {cell's} surface area, $\varepsilon$ is the emissivity of the {cell's} material, and $T_0$ its temperature. The absorbed radiance is the fraction of incident radiance that is not reflected (cf. \eqref{eq:emiss}). The incoming radiance is modeled with
\begin{equation}
 E_\mathrm{in}=\sigma\cdot T_\mathrm{amb}^4,  
 \label{eq:incidentRadiance}
\end{equation}
where $T_\mathrm{amb}$ is the ambient temperature that we assume to be constant. 
To model $p_\mathrm{rad}$, we simplify \eqref{eq:radPower} and \eqref{eq:incidentRadiance} to {
\begin{equation}
p_\mathrm{rad} = r\cdot \varepsilon\cdot \left( T_0^4 - T_\mathrm{amb}^4\right),
\label{eq:radPowerModel}
\end{equation}}
where the parameter $r$ is determined by training. 

The four powers $p_\mathrm{T}$, $p_\mathrm{B}$, $p_\mathrm{L}$, and $p_\mathrm{R}$ represent the power flows to adjacent {cells} as shown in Fig.~\ref{fig:thermResis} and are given by \cite{Paek13}
\begin{equation}
p_d = C_d\cdot\left(T_d - T_0\right), 
\label{eq:gradient_flow}
\end{equation}
where the direction  $d\in\{\mathrm{T}, \mathrm{B}, \mathrm{L}, \mathrm{R}\}$ refers to the four adjacent {cells}, $C_d$ is the thermal conductance between the two {cells}, $T_d$ the temperature of the adjacent {cell}, and $T_0$ the temperature of the current {cell}. 

Finally, $p_\mathrm{c}$ is the heat flux to the ambient air as \cite{Penoncello19}
\begin{equation}
p_\mathrm{c} = h\cdot\left( T_\mathrm{amb} - T_0 \right), 
\label{eq:heatFlux}
\end{equation}
where $T_\mathrm{amb}$ is the ambient room temperature, which is again assumed to be constant. Moreover, $h$ is the convective heat transfer coefficient. 

These equations are transferred to the 2D space, which corresponds to the thermal image space, and summarized in Algorithm\,\ref{alg:t2p}. In this algorithm, the operator ``$* *$'' is a two-dimensional convolution and ``$\circ$'' the Hadamard-product, which is an element-wise multiplication. The gradient filters are defined as 
\begin{equation}
\boldsymbol{G}_\mathrm{h} =\begin{pmatrix}
1 \\ -1 \\
\end{pmatrix}, \quad
\boldsymbol{G}_\mathrm{v} =\begin{pmatrix}
1 & -1 \\
\end{pmatrix}.
\end{equation}

\begin{algorithm}[h!]
\label{alg:t2p}
\providecommand\comment{\color[rgb]{0.35,0.35,0.35}}
\SetKwInOut{Input}{input}\SetKwInOut{Output}{output}
\Input{$\boldsymbol{T}_\mathrm{in}$, $T_\mathrm{amb}$, $\mathcal{M}$, $\mathcal{C}$, $ {\mathcal{E}}$, $h$, $r$}
\Output{ 
$\boldsymbol{P}_\mathrm{el}$;{\comment{ \slash\slash ~Pixel-area-wise electrical power }} \\}
{
\For {$h \gets \{2, 3, ...H-1\}$}{
\For {$w \gets \{2, 3, ...W-1\}$}{ 
	{\comment{ \slash\slash ~Get emissivities and inverse emissivities for all pixels using indexing in $\mathcal{E}$}:} \\
	 { $\boldsymbol{O}(h,w)\gets \mathcal{E}\Big( \mathcal{M}(h,w)   \Big)$;}\\
	 { $\boldsymbol{O}_\mathrm{inv}(h,w)\gets \frac{1}{\boldsymbol{O}(h,w)}$;}\\
	{\comment{ \slash\slash ~Get thermal conductance matrices for all directions using indexing in $\mathcal{C}$ \eqref{eq:symmetricConductanceMatrix}}:} \\
	$\boldsymbol{C}_\mathrm{T}(h,w)\gets \mathcal{C}\Big( \mathcal{M}(h,w), \mathcal{M}(h-1,w)   \Big)$;\\
	$\boldsymbol{C}_\mathrm{B}(h,w)\gets \mathcal{C}\Big( \mathcal{M}(h,w), \mathcal{M}(h+1,w)   \Big)$;\\
	$\boldsymbol{C}_\mathrm{L}(h,w)\gets \mathcal{C}\Big( \mathcal{M}(h,w), \mathcal{M}(h,w-1)   \Big)$;\\
	$\boldsymbol{C}_\mathrm{R}(h,w)\gets \mathcal{C}\Big( \mathcal{M}(h,w), \mathcal{M}(h,w+1)   \Big)$; \\
	}
}	
{ 
{\comment{ \slash\slash ~Compensate emissivity: }}\\
$\boldsymbol{T}_\mathrm{em} \gets \sqrt[4]{\left(\boldsymbol{T}_\mathrm{in}^4-\left(1-\boldsymbol{O}\right) \cdot T_\mathrm{amb}\right)\circ \boldsymbol{O}_\mathrm{inv}}$;} \\
\comment{\slash\slash ~Get spatial temperature gradients \eqref{eq:gradient_flow}: }}\\
	$\Delta\boldsymbol{T}_\mathrm{T} \gets \boldsymbol{T}_\mathrm{em} * *\ \boldsymbol{G}_\mathrm{h} $;\quad
	$\Delta\boldsymbol{T}_\mathrm{B} \gets \boldsymbol{T}_\mathrm{em} * * \left(-\boldsymbol{G}_\mathrm{h}\right) $; \\
	$\Delta\boldsymbol{T}_\mathrm{L} \gets \boldsymbol{T}_\mathrm{em} * *\ \boldsymbol{G}_\mathrm{v} $;\quad
	$\Delta\boldsymbol{T}_\mathrm{R} \gets \boldsymbol{T}_\mathrm{em} * * \left(-\boldsymbol{G}_\mathrm{v}\right) $;\\
$\Delta\boldsymbol{T}_\mathrm{amb} \gets T_\mathrm{amb}-\boldsymbol{T}_\mathrm{em}$;{\comment{ \slash\slash ~Temperature difference to ambient temperature }} \\
$\Delta\boldsymbol{T}_\mathrm{amb}^4 \gets T_\mathrm{amb}^4-\boldsymbol{T}_\mathrm{em}^4$;{\comment{ \slash\slash ~Fourth power temperature difference. }} \\
{\comment{ \slash\slash ~Calculate power flow on the board \eqref{eq:gradient_flow}: }}\\
$\boldsymbol{P}_\mathrm{T}\gets \boldsymbol{C}_\mathrm{T}\circ \Delta\boldsymbol{T}_\mathrm{T} $; \quad
$\boldsymbol{P}_\mathrm{B}\gets \boldsymbol{C}_\mathrm{B}\circ \Delta\boldsymbol{T}_\mathrm{B} $; \\
$\boldsymbol{P}_\mathrm{L}\gets \boldsymbol{C}_\mathrm{L}\circ \Delta\boldsymbol{T}_\mathrm{L} $; \quad\
$\boldsymbol{P}_\mathrm{R}\gets \boldsymbol{C}_\mathrm{R}\circ \Delta\boldsymbol{T}_\mathrm{R} $;\\

$\boldsymbol{P}_\mathrm{c}\gets h\cdot \Delta\boldsymbol{T}_\mathrm{amb}$;{\comment{ \slash\slash ~Convective power flow \eqref{eq:heatFlux} }} \\

$\boldsymbol{P}_\mathrm{rad}\gets r\cdot {\boldsymbol{O} \circ }\Delta\boldsymbol{T}_\mathrm{amb}^4$;{\comment{ \slash\slash ~Radiative power flow \eqref{eq:radPowerModel} }} \\

$\boldsymbol{P}_\mathrm{el} \gets -\boldsymbol{P}_\mathrm{T}-\boldsymbol{P}_\mathrm{B}-\boldsymbol{P}_\mathrm{L}-\boldsymbol{P}_\mathrm{R}-\boldsymbol{P}_\mathrm{c}-\boldsymbol{P}_\mathrm{rad}$;{\comment{ \slash\slash ~Power sum \eqref{eq:pixelModel} }} \\
\caption{Temperature-to-power conversion algorithm. The input comprises the temperature map $\boldsymbol{T}_\mathrm{in}$, the ambient temperature $T_\mathrm{amb}$, the component map $\mathcal{M}$, the matrix {of heat conductivities $\mathcal{C}$, the emissivities $\mathcal{E}$}, the convective heat transfer coefficient $h$, and the radiance coefficient $r$.  The output is an estimation of the electrically dissipated power $\boldsymbol{P}_\mathrm{el}$ for each {cell}. 
}
\end{algorithm}

{
The algorithmic flow of the power estimation in Algorithm~\ref{alg:t2p} works as follows. First (lines 1-12), the emissivities from $\mathcal{E}$ and the heat transfer coefficients from $\mathcal{C}$ are mapped onto 2D arrays of the input image size, such that each entry gets assigned the values corresponding to the material of the corresponding cell at the position of the input temperature map. These arrays can then be used for matrix operations with the input temperature map $\boldsymbol{T}_\mathrm{in}$. Then, the emissivity compensation \eqref{eq:temp_modify} is performed to obtain more realistic temperature values (line 14). Afterwards, in lines 16-17, the temperature gradients on the 2D surface (cell-to-cell difference) are computed. Subsequently, in lines 18-19, the temperature difference of all cells with respect to the ambient temperature is calculated, where the linear difference is used to determine the convective heat loss \eqref{eq:heatFlux} and the difference to the power of four is used for the radiative heat loss \eqref{eq:radPowerModel}. In lines 21-22, the power flow due to heat conductance on the board is determined \eqref{eq:gradient_flow}. In contrast to the SIMP approach presented in \cite{Hamann07}, where only the heat flow due to thermal conductance was considered, we also take into account the radiative power flow \eqref{eq:radPowerModel} and the convective power flow \eqref{eq:heatFlux} in lines 23-24. Finally, in line 25, the cell-wise power flow in the steady state is calculated by adding all power flow components \eqref{eq:pixelModel}.
}

\subsection{Ambient Temperature}
\label{secsec:ambTemp}
In order to mathematically describe the convective heat flux as shown in \cite{Penoncello19}, the difference between the body's temperature and the ambient temperature $T_\mathrm{amb}$ is needed as shown in \eqref{eq:heatFlux}. To estimate $T_\mathrm{amb}$, we propose to exploit the information that is already available in the temperature map $\boldsymbol{T}_\mathrm{in}$, which is recorded by the thermal camera. To this end, we assume that most parts of the map represent temperatures that are equal or close to the ambient temperature, because active, hot components are only located in the center of the picture.  
\begin{figure}[tb]
\begin{minipage}[b]{.26\linewidth}
  \centering
  \centerline{\includegraphics[width=.95\textwidth]{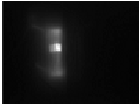}}
  \centerline{(a)}
\end{minipage}
\begin{minipage}[b]{0.18\linewidth}
   \psfrag{0}[l][l]{$25^\circ$C }
	\psfrag{1}[l][l]{$75^\circ$C }
	\psfrag{2}[l][l]{$125^\circ$C }
  \centerline{\hspace{-.8cm}\includegraphics[width=.45\textwidth]{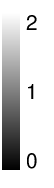}}
  \vspace{-.5cm}
  \centerline{}\medskip
\end{minipage}
\begin{minipage}[b]{0.55\linewidth}
  \centering
\psfrag{000}[ct][ct]{ }
\psfrag{001}[ct][ct]{ $40$}%
\psfrag{002}[ct][ct]{ }
\psfrag{003}[ct][ct]{ $80$}%
\psfrag{004}[ct][ct]{ }
\psfrag{005}[ct][ct]{ $120$}%
\psfrag{006}[ct][ct]{ }
\psfrag{007}[rc][rc]{ $1$}%
\psfrag{008}[rc][rc]{ $10^2$}%
\psfrag{009}[rc][rc]{ $10^4$}%
\psfrag{010}[t][b]{ Temperature [$^\circ$C]}%
\psfrag{011}[b][t]{ Freq. of Occ.}%
  \centerline{\includegraphics[width=.95\textwidth]{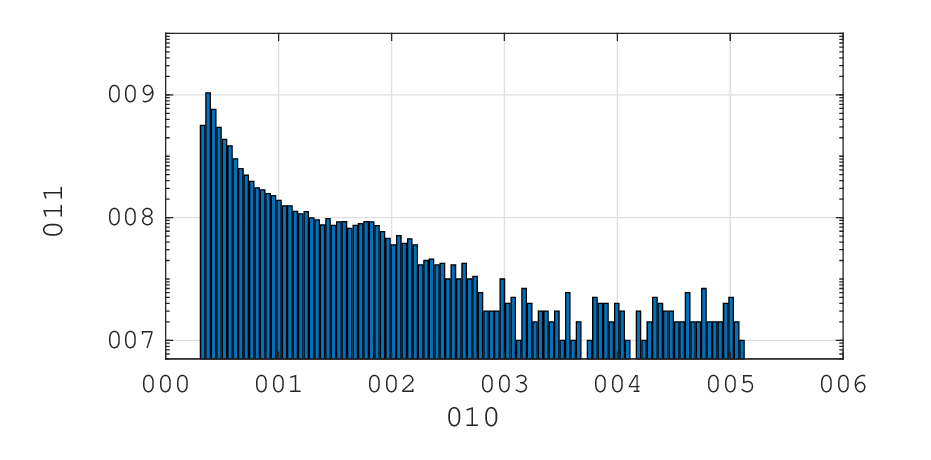}}
  \centerline{(b)}
\end{minipage}
\caption{Distribution of cell-wise temperature values for a temperature map with active components. Fig.~(a) shows an example for the IR-sensor data, i.e. the thermal map for four active resistors. Bright pixels represent high temperatures and dark pixels represent cold temperatures. Fig.~(b) shows a histogram of the temperature distribution. 
The maximum can be found close to $27^\circ$C. }
\label{fig:thermalHist}
\end{figure}

Figure~\ref{fig:thermalHist} (a) shows a temperature map with four active resistors in the center. The picture indicates that increased temperatures only occur close to these resistors, whereas at the borders, the temperatures are approximately constant. For the proposed algorithm, we assume that these temperatures are representative for the ambient temperature $T_\mathrm{amb}$. 

To extract this ambient temperature, we first calculate a histogram with $N=100$ bins as shown in Fig.~\ref{fig:thermalHist} (b). We find that a temperature value close to $27^\circ$C occurs most frequently. 
Then, we take the center temperatures of the maximum bin as well as the two adjacent bins (the bins to the left and to the right) and weight the temperatures with the corresponding frequencies of occurrences. The resulting weighted mean yields the ambient temperature $T_\mathrm{amb}$ used in \eqref{eq:heatFlux}. {Note that we do not perform emissivity compensation for the ambient temperature because we assume that the material of the corresponding cells has an emissivity close to one. }

\begin{figure*}[ht]
\psfrag{T}[c][c]{$\boldsymbol{T}_\mathrm{in}$}
\psfrag{N}[l][l]{$\mathcal{M}$}
\psfrag{Z}[l][l]{Model parameters}
\psfrag{c}[c][t]{$\mathcal{C}$}
\psfrag{m}[c][t]{$\mathcal{E}$}
\psfrag{h}[c][c]{$h,r$}
\psfrag{i}[l][l]{$\boldsymbol{T}_\mathrm{em}$}
\psfrag{k}[l][l]{$\boldsymbol{C}_i$}
\psfrag{a}[c][c]{$T_\mathrm{amb}$}
\psfrag{D}[l][l]{$\Delta\boldsymbol{T}_i$}
\psfrag{F}[c][c]{$\boldsymbol{P}_\mathrm{el}$}
\psfrag{H}[c][c]{$\hat{ {P}}_k$}
\psfrag{E}[c][c]{\small{Emissivity}}
\psfrag{C}[c][c]{\small{compensation}}
\psfrag{A}[c][c]{\small{Ambient}}
\psfrag{t}[c][c]{\small{temperature}}
\psfrag{z}[c][c]{\small{Temperature}}
\psfrag{d}[c][c]{\small{determination}}
\psfrag{R}[c][c]{\small{Conductance}}
\psfrag{M}[c][c]{\small{matrix}}
\psfrag{G}[c][c]{\small{generation}}
\psfrag{g}[c][c]{\small{gradients}}
\psfrag{x}[c][c]{\small{Pixel-wise}}
\psfrag{w}[c][cb]{\small{power}}
\psfrag{e}[c][c]{\small{estimation}}
\psfrag{K}[c][c]{\small{Component-wise}}
\psfrag{b}[c][c]{\small{aggregation}}
  \centering
	\includegraphics[width=.8\textwidth]{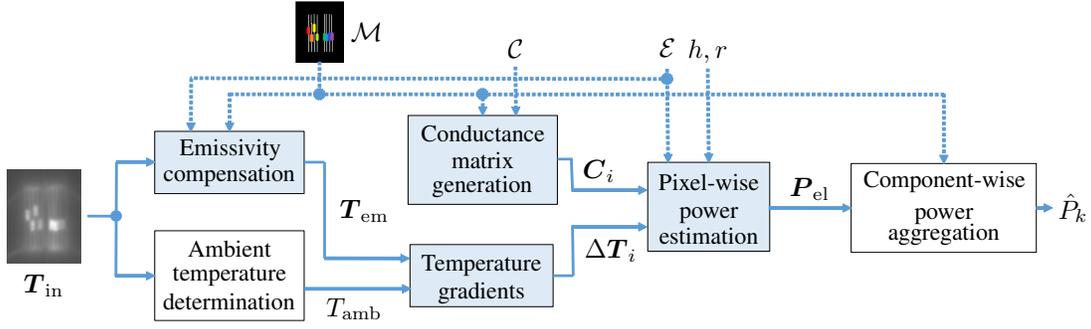}
	\caption{Algorithmic flow-graph of the component-wise power estimation algorithm. The observed temperature map from the thermal camera $\boldsymbol{T}_\mathrm{in}$ is the input. As an output, the component-wise powers $\hat{{P}}_k$ for each component $k$ are returned. }
\label{fig:algoFlow}
\end{figure*}

\subsection{Component-wise Power Aggregation}
\label{secsec:compPower}
Using Algorithm\,\ref{alg:t2p}, we can calculate estimates of the electrical power dissipation {for each cell}. However, in this paper, we are interested in the estimation of the overall power of a predefined set of components on the board such as a chip or a resistor. As such, we sum up the estimated dissipated power from each {cell} corresponding to the desired component $k$ as 
\begin{equation}
\hat P_k = \sum_{(m,n)} \boldsymbol{P}_\mathrm{el} \circ \mathcal{M}_k, 
\label{eq:compPower}
\end{equation}
where $\mathcal{M}_k$ is a binary mask extracted from the component map $\mathcal{M}$, whose entries are $1$ if the {cell} corresponds to the desired component $k$ and $0$ otherwise. The circumflex on  $\hat P_k$ indicates that we refer to the estimated power of the $k$-th component.

A flow graph of the component-wise power estimation algorithm is illustrated in Fig.~\ref{fig:algoFlow}. 
 The emissivity compensation was explained in Subsection~\ref{secsec:emiss}, ambient temperature determination in Subsection~\ref{secsec:ambTemp}, and the calculations from Algorithm~\ref{alg:t2p}  in Subsection~\ref{secsec:pixModel} are represented by the blue-colored boxes. 

\section{Measurement Setup}
\label{sec:setup}

This section explains the details on the measurement setup that was developed to provide a proof of concept of the proposed component-wise power estimation method. Subsection~\ref{secsec:hardware} presents the hardware setup including the carrier board, the resistors, and the thermal camera. Then, Subsection~\ref{secsec:visual} shows thermal image examples for a certain setup and a corresponding component map. Finally, Subsection~\ref{secsec:measurementSet} explains the set of measurements with different resistor layouts. 

\subsection{Hardware Setup}
\label{secsec:hardware}
\begin{figure}[ht]
\begin{minipage}[b]{.48\linewidth}
  \centering
  \centerline{\includegraphics[width=3.35cm]{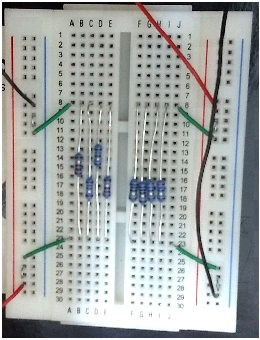}}
  \centerline{(a) Board with resistors}\medskip
\end{minipage}
\hfill
\begin{minipage}[b]{0.48\linewidth}
  \centering
  \centerline{\includegraphics[width=4.0cm]{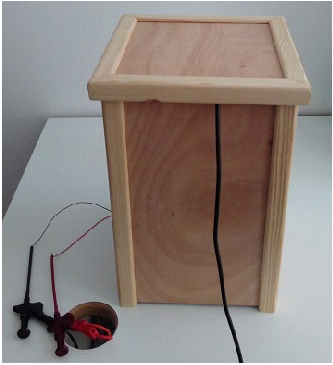}}
  \centerline{(b) Wooden box}\medskip
\end{minipage}
\caption{Photographs of the measurement setup. (a) Carrier board with eight different resistors. (b) Wooden box for controlling the incident radiation. }
\label{fig:photo}
\end{figure}

As an electronic circuit board, we use a carrier board with resistors as shown in Fig.~\ref{fig:photo} (a). This board provides the advantage that it is freely configurable and power dissipation $P_R$ of each resistor is known due to Ohm's law
\begin{equation}
P_R = \frac{V^2}{R}, 
\label{eq:Ohm}
\end{equation}
where $V$ is the input voltage and $R$ is the resistance. For accurate power values, a resistor tolerance of $1\%$ is chosen. 

We put a wooden box around the camera and the board as shown in Fig.~\ref{fig:photo}~(b). We assume that the ambient temperature and the temperature of the wooden box is constant such that all incident radiation to the board $E_\mathrm{box}$ in \eqref{eq:emiss} is also constant and solely determined by the ambient temperature $T_\mathrm{amb}$. 

The IR-camera is a consumer device for portable applications with a resolution of $W=206$ pixels and $H=156$ pixels. The sensor is a microbolometer  with a $36^\circ$ field-of-view, a frame rate of up to $9$\,Hz and a temperature range of $-40^\circ$C to $3360^\circ$C. The microbolometer uses vanadium oxide as sensor material. The image data used for modeling is the raw temperature data obtained through a USB connection.

\subsection{Visual Data}
\label{secsec:visual}
\begin{figure}[tb]
\begin{minipage}[b]{.26\linewidth}
  \centering
  \centerline{\includegraphics[width=.95\textwidth]{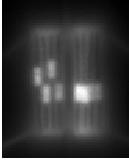}}
  \centerline{(a)}
\end{minipage}
\begin{minipage}[b]{0.26\linewidth}
  \centering
  \centerline{\includegraphics[width=.95\textwidth]{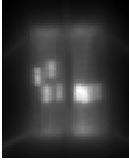}}
  \centerline{(b)}
\end{minipage}
\begin{minipage}[b]{0.18\linewidth}
   \psfrag{0}[l][l]{$25^\circ$C }
	\psfrag{1}[l][l]{$75^\circ$C }
	\psfrag{2}[l][l]{$125^\circ$C }
  \centerline{\hspace{-.8cm}\includegraphics[width=.45\textwidth]{gfx/colormapping}}
  \centerline{}\medskip
\end{minipage}
\begin{minipage}[b]{0.26\linewidth}
  \centering
  \centerline{\includegraphics[width=.95\textwidth]{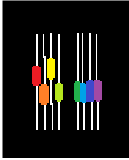}}
  \centerline{(c)}
\end{minipage}
\caption{Visualization of the input data for power estimation. (a) Example for the IR-sensor data corresponding to the board in Fig.~\ref{fig:photo}, where bright represents high temperatures and dark represents cold temperatures. (b) Estimated temperature after bicubic interpolation for metal wires. (c) Visualization of the component map $\mathcal{M}$, where black represents the background (i.e. the board), white represents metal wires, and the colors indicate the position of the resistors. }
\label{fig:thermalMaps}
\end{figure}

The sensed temperature data from the  thermal IR-camera is visualized in Fig.~\ref{fig:thermalMaps}. 
Fig.~\ref{fig:thermalMaps}~(a) shows the raw temperature data for a constant voltage over all resistors in the steady state (approx. two minutes after switching on). We can see that the resistors show the highest temperature and that the heat is distributed to the top and to the bottom through the metal wires. Left and right of the resistors, the temperature gradient is steeper. 

Closely investigating the grey levels at the positions of the wires in Fig.~\ref{fig:thermalMaps} (a), one can see that the sensed temperature of the wires is smaller than the surrounding temperatures corresponding to the carrier board. This is indicated by dark vertical lines above and below the resistors. These dark lines are caused by the low emissivity of the wires, which reflect the incident radiation $E_\mathrm{in} = E_\mathrm{box}$ caused by the wooden box. The magnitude of this radiation depends on the ambient temperature, which is smaller than the temperature of the wires, thus causing the wires to appear darker, which does not reflect the true temperature. 

Figure~\ref{fig:thermalMaps} (b) illustrates the proposed solution of bicubic interpolation (cf. Section~\ref{secsec:emiss}). Interpolation is performed on all pixels corresponding to wires as well as pixels adjacent to wires. The pixels corresponding to wires are visualized in Fig.~\ref{fig:thermalMaps} (c) by the white, vertical lines. 

Furthermore, Fig.~\ref{fig:thermalMaps} (c) visualizes the positions of the resistors, i.e. the components indexed by $k$ based on the component map $\mathcal{M}$. Each color refers to a single resistor. For each resistor configuration, such a map has been produced by manual labeling.

\subsection{Set of Measurements}
\label{secsec:measurementSet}
We measure multiple configurations of resistors. The configurations vary with the number, the positions, and the resistance values of the resistors. Table~\ref{tab:configs} lists the configurations that were tested, where configuration B corresponds to the example shown in Fig.~\ref{fig:thermalMaps}. 

\begin{table}[t]
\renewcommand{\arraystretch}{1.3}
\caption{Resistor configurations for evaluating the power estimation algorithm. The resistor index is used in the component map $\mathcal{M}$ and visualized for configuration B with different colors in Fig.~\ref{fig:thermalMaps}~(c). }
\label{tab:configs}
\begin{center}
\begin{tabular}{r||r|r|r|r}
\hline
  &\multicolumn{4}{c}{Configuration}\\
Resistor Index $k$ & A & B & C & D \\
\hline
$1$&$27\ \Omega$&$12\ \Omega$& $1000\ \Omega$ & $150\ \Omega$\\
$2$&$27\ \Omega$&$15\ \Omega$& $1000\ \Omega$ & $150\ \Omega$\\
$3$&$10\ \Omega$&$18\ \Omega$& $100\ \Omega$ & $220\ \Omega$\\
$4$&$10\ \Omega$&$22\ \Omega$& $100\ \Omega$ & $220\ \Omega$\\
$5$&-&$10\ \Omega$& -&-\\
$6$&-&$10\ \Omega$& -&-\\
$7$&-&$27\ \Omega$& -&-\\
$8$&-&$27\ \Omega$& -&-\\
\hline
 \end{tabular}
\end{center}
\end{table}

We apply $10$ different input voltages to the resistor configurations to simulate different amounts of dissipated powers resulting in $200$ independent measurement instances. The true power dissipation of the resistors is calculated using \eqref{eq:Ohm}. To make sure that the thermal data is recorded in the steady state, we wait for two minutes after switching on the supply voltage before reading the thermal images from the IR-camera. Furthermore, we take five readings for each configuration and each input voltage in order to allow the calculation of the measurement uncertainty such that in total, $1000$ readings were performed. 
Throughout all configurations, dissipated powers of the resistors span from $0.9$\,mW to $1$\,W.

\section{Evaluation}
\label{sec:eval}
{ We evaluate and compare the performance of the proposed algorithm using different variants. At first, we test the estimation error of the full algorithm presented in Algorithm~\ref{alg:t2p} (FULL). Second, we adopt the second approach from Subsection~\ref{secsec:emiss}, in which the temperature of materials with a low emissivity is interpolated instead of compensated (INT). In this case, the equation in line 14 of Algorithm~\ref{alg:t2p} is simplified to $\boldsymbol{T}_\mathrm{em} \gets \boldsymbol{T}_\mathrm{in}$.  Interpolation of input temperature values $\boldsymbol{T}_\mathrm{in}$ is performed on pixels indicated as wires and their neighboring pixels. }

{ Third, on top of the simplification mentioned above, we test the influence of the radiative power on the modeling accuracy by removing \eqref{eq:radPowerModel} from the power flow model \eqref{eq:pixelModel} with $\boldsymbol{P}_\mathrm{rad}\gets 0$ in line 24 of Algorithms~\ref{alg:t2p} (NORAD). Finally, we additionally remove the influence of the heat flux by removing \eqref{eq:heatFlux} with $\boldsymbol{P}_\mathrm{c}\gets 0$ in l. 23 of Algorithms~\ref{alg:t2p} (NOFLUX), which is the closest solution to the approach presented by Hamann et al. \cite{Hamann07}, where only the power flow due to horizontal heat conductivity is considered.  }

We train the respective parameters $\mathcal{C}$, $\mathcal{E}$, $h$, and $r$ { for each configuration} using least-squares curve fitting by exploiting a trust-region reflective algorithm \cite{Coleman96}.
The minimization criterion is defined as 
\begin{align}
&\min_{\mathcal{C},h,r} \tilde e =   \\ 
& \notag  \min_{\mathcal{C},h,r} \left( \sum_{l,i,k} \left(\hat P_{k,i,l} - P_{k,i,l}\right)^2 + \hat P_{\mathrm{board},i}^2 + \hat P_{\mathrm{wire},i}^2\right) ,
\end{align}
where $\tilde e$ is the error to be minimized. 
 The sum represents the overall squared error over all voltages $V_i$, all resistors $k$, and all configurations $l$, where $\hat P_{k,i,l} $ is the estimated power from \eqref{eq:compPower} and $P_{k,i,l} $ the true power. The latter two summands are calculated by 
\begin{align}
\hat P_{\mathrm{board}}^2 = \sum_{(m,n)} \left(\boldsymbol{P}_\mathrm{el} \circ \mathcal{M}_\mathrm{board}\right)^2
\label{eq:compPowerBoard} \\
\hat P_{\mathrm{wire}}^2 = \sum_{(m,n)} \left(\boldsymbol{P}_\mathrm{el} \circ \mathcal{M}_\mathrm{wire}\right)^2, 
\label{eq:compPowerWire} 
\end{align}
where $\mathcal{M}_\mathrm{board}$ and $\mathcal{M}_\mathrm{wire}$ are the binary pixel masks for the board (i.e. the background) and the metal wires, respectively. The squaring operation is performed pixel-wise. The terms $\hat P_{\mathrm{board}}^2$ and $\hat P_{\mathrm{wire}}^2$ in the error function can be interpreted as regularization terms, which ensure that the modeled dissipated power of inactive components is close to zero. { As the number of parameters to be trained is  $11$ maximum (see Table~\ref{tab:stdErrorsAllAlgos}), we also need at least $11$ discrete measurements for a unique solution during training. To still allow proper evaluation, }we perform a $10$-fold cross-validation to separate the training set from the validation set{, where in each iteration, we have $36$ samples for training and $4$ samples for validation. }

\subsection{Comparison}
{
Before providing a detailed discussion on measurement results, we perform a comparison of the four algorithms described in the last subsection. Table~\ref{tab:stdErrorsAllAlgos} lists the variants, their corresponding number of parameters to be trained, and the standard deviation of the estimated powers with respect to the true power given as 
\begin{equation}
e_\mathrm{std} = \sqrt{\frac{1}{\left|L\right|}\sum_{l\in L}\frac{1}{I_l K_l}\sum_{i=1}^{I_l}\sum_{k=1}^{K_l} \left(\hat P_{k,i,l} - P_{k,i,l}\right)^2}, 
\label{eq:absErr}
\end{equation}
where $\left|L\right|=4$ is the number of configurations and $l\in L=\{\mathrm{A},\mathrm{B},\mathrm{C},\mathrm{D}\}$ the configuration index; $K_l$ is the number of resistors of configuration $l$ and $I_l=10$ is the number of measurements, i.e. the number of different input voltages. 
\begin{table}[t] 
\renewcommand{\arraystretch}{1.3}
\caption{{Number of training parameters and estimation error in terms of standard deviation depending on the algorithm.} }
\vspace{-0.5cm}
\label{tab:stdErrorsAllAlgos}
\begin{center}
\begin{tabular}{r||c|c|c|c}
\hline
Algorithm & FULL & INT & NORAD & NOFLUX \\
\hline
No. params. & $11$ & $8$ & $7$ & $6$ \\
$e_\mathrm{std} $ &$40.86$\,mW & $40.68$\,mW &$41.94$\,mW & $69.74$\,mW\\
\hline
\end{tabular}
\end{center}
\end{table}

The results indicate that the INT method returns the best estimation accuracy (standard deviation of $40.68$\,mW). This means that interpolation of uncertain temperatures, as obtained by power readings from the wires having a low emissivity, is more accurate and stable than a direct compensation of the emissivity using a training process (FULL). 

Furthermore, we can see that neglecting the radiative power (NORAD) and the heat flux (NOFLUX), the estimation error rises to $41.94$\,mW and $69.74$\,mW, respectively. Apparently, the radiative power has a small influence on the estimation error, but neglecting the heat flux leads to severe degradations in modeling accuracy. In the following, the INT algorithm will be evaluated in more detail as it returns the highest estimation accuracy. 
}

\subsection{Visual Results}
\begin{figure}[tb]
\centering
\begin{minipage}[b]{.3\linewidth}
  \centering
  \centerline{\includegraphics[width=.95\textwidth]{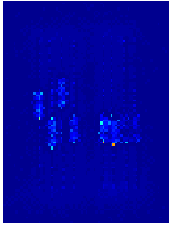}}
  \centerline{(a) $V=1.25$\,V}
\end{minipage}
\begin{minipage}[b]{0.3\linewidth}
  \centering
  \centerline{\includegraphics[width=.95\textwidth]{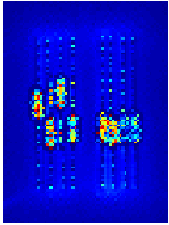}}
  \centerline{(b) $V=2.25$\,V}
\end{minipage}
\begin{minipage}[b]{0.2\linewidth}
   \psfrag{0}[l][l]{$0$\,$\frac{\mathrm{mW} }{\mathrm{pixel}}$ }
	\psfrag{1}[l][l]{$10$\,$\frac{\mathrm{mW} }{\mathrm{pixel}}$ }
	\psfrag{2}[l][l]{$20$\,$\frac{\mathrm{mW} }{\mathrm{pixel}}$ }
  \centerline{\hspace{-.8cm}\includegraphics[width=.45\textwidth]{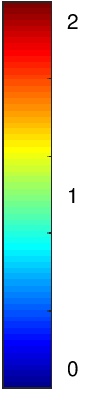}}
  \centerline{}\medskip
\end{minipage}
\caption{Estimated power maps for configuration B and $V=1.25$~V (a) and $V=2.25$~V (b). The pixel color is the estimated power per pixel. 
}
\label{fig:estPowerMap}
\end{figure}

After training the parameters $\mathcal{C}$, $h$, and $r$ { for the INT algorithm}, the power model from Algorithm~\ref{alg:t2p} can be applied to the input temperature data. The corresponding visual results for configuration B with voltages $V=1.25$\,V and $V=2.25$\,V are illustrated in Fig.~\ref{fig:estPowerMap} (a) and (b), respectively. 

The power maps show that the dissipating components, i.e., the resistors, are detected at the correct positions. In addition, highest per-pixel power values are detected at the resistors that have the highest power dissipation in the test setup. However, it can be seen that the variation of the per-pixel power is very high, which can be attributed to measurement noise, which is mainly caused by the consumer thermal camera.

\subsection{Estimated Powers}
Exemplary values for the estimated and the true power dissipation are shown in Table~\ref{tab:cpPowers}. The values correspond to the power values for resistors $k=2$ and $k=8$ of Configuration B. 
\begin{table}[t] 
\renewcommand{\arraystretch}{1.3}
\caption{Estimated and true power values in mW for resistors $k=2$ and $k=8$ of Configuration~B.  }
\vspace{-0.5cm}
\label{tab:cpPowers}
\begin{center}
\begin{tabular}{r||r|r||r|r}
\hline
$V_i$\,[V] & $P_{k=2,i,B}$ &  $\hat P_{k=2,i,B}$ & $P_{k=8,i,B}$&$\hat P_{k=8,i,B}$\\
\hline
$0.25$ &  $3.69$ & $3.94$&$2.05$& $-0.35$\\
$0.50$ &  $14.8$ & $12.8$&$8.20$&$3.27$\\ 
$0.75$ &  $33.2$&$26.7$& $18.4$&$12.5$ \\
$1.00$ &  $58.9$& $49.5$&$32.7$&$24.4$ \\
$1.25$ & $92.2$&$83.9$&  $51.2$& $43.3$\\
 $1.50$ &  $132$&$123$&$73.5$&$65.1$\\
 $ 1.75$ & $180$&$176$&$100$&$93.9$\\
 $ 2.00$ & $235$&$241$&$131$&$129$\\
 $ 2.25$ & $298$&$311$&$165$&$165$\\
 $2.50$ & $368$&$383$&$204$ &$200$\\ 
 \hline
   \end{tabular}
\end{center}
\end{table}
The table shows that first, the estimates generally follow the trend of the true power values. Second, it is striking that the estimated value can be negative {(top row, right)}, which can be attributed to measurement noise. In this case, the negative value occurs because the measured temperatures on the resistor are slightly smaller than the ambient temperature.

\subsection{Measurement Uncertainty}
In this section, we evaluate the measurement uncertainty of the proposed method by analyzing the spread of the estimated power values and the estimation error { for the INT algorithm}. To determine the spread, we first calculate the estimated powers for all $1000$ readings. Afterwards, we calculate the standard deviation of the powers \eqref{eq:absErr} over the five readings of each measurement instance, such that we obtain $200$ values describing the spread. The maximum spread is found to be $14\,\mathrm{mW}$ and the mean spread of all $200$ instances yields $1.2\,\mathrm{mW}$. 

Furthermore, we consider the estimation error for the power dissipation of all resistors by calculating the standard deviation \eqref{eq:absErr}
and the mean relative error
\begin{equation}
e_\mathrm{rel} = \frac{1}{\left|L\right|}\sum_{l\in L}\frac{1}{I_l K_l}\sum_{i=1}^{I_l}\sum_{k=1}^{K_l} \left|\frac{\hat P_{k,i,l} - P_{k,i,l}}{P_{k,i,l}}\right|, 
\label{eq:relErr}
\end{equation}
respectively. { As shown in Table~\ref{tab:stdErrorsAllAlgos}, the estimation error is $e_\mathrm{std}=40.68\,\mathrm{mW}$,} which is significantly larger than the spread. This error shows that in particular, low power values ($<100\,\mathrm{mW}$) cannot be estimated accurately. {The mean relative error is found to be $35\%$}.

\begin{figure}[t]
\providecommand\matlabtextB{\color[rgb]{0.850,0.325,0.098}}
\providecommand\matlabtextL{\color[rgb]{0.0,0.447,0.741}}
\psfrag{014}[bc][bc]{ Mean relative error $e_\mathrm{rel}$~[\%]}%
\psfrag{015}[tc][tc]{\matlabtextB Standard deviation $e_\mathrm{std}$ [mW]}%
\psfrag{013}[tc][tc]{ Minimum considered power $P_\mathrm{min}$~[mW]}%
\providecommand\matlabtextD{\color[rgb]{0.150,0.150,0.150}}%
\psfrag{000}[ct][ct]{\matlabtextD $0$}%
\psfrag{001}[ct][ct]{\matlabtextD $100$}%
\psfrag{002}[ct][ct]{\matlabtextD $200$}%
\psfrag{003}[ct][ct]{\matlabtextD $300$}%
\psfrag{004}[ct][ct]{\matlabtextD $400$}%
\psfrag{005}[ct][ct]{\matlabtextD $500$}%
\providecommand\matlabtextE{\color[rgb]{0.000,0.000,0.000}}%
\psfrag{35}[rc][rc]{ $35$}%
\psfrag{5}[rc][rc]{ $5$}%
\psfrag{10}[rc][rc]{ $10$}%
\psfrag{15}[rc][rc]{ $15$}%
\psfrag{20}[rc][rc]{ $20$}%
\psfrag{25}[rc][rc]{ $25$}%
\psfrag{30}[rc][rc]{ $30$}%
\psfrag{0.01}[lc][lc]{\matlabtextB $0.01$}%
\psfrag{0}[lc][lc]{\matlabtextB $0$}%
\psfrag{006}[lc][lc]{\matlabtextB $40$}%
\psfrag{007}[lc][lc]{\matlabtextB $50$}%
\psfrag{008}[lc][lc]{\matlabtextB $60$}%
\psfrag{009}[lc][lc]{\matlabtextB $70$}%
\psfrag{010}[lc][lc]{\matlabtextB $80$}%
\psfrag{011}[lc][lc]{\matlabtextB $90$}%
\psfrag{012}[lc][lc]{\matlabtextB $100$}%
  \centering
	\includegraphics[width=.48\textwidth]{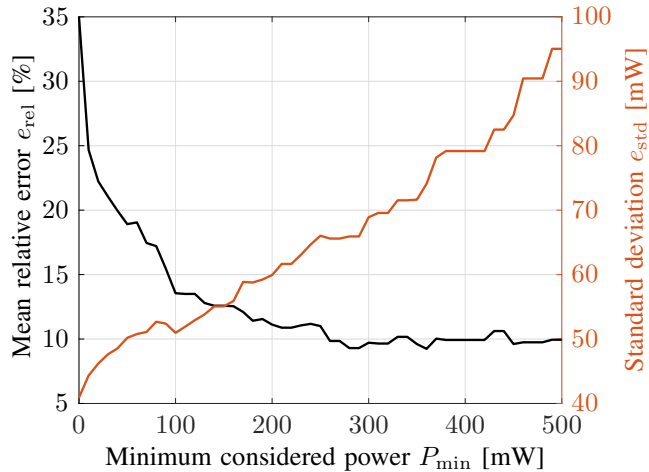}
	\caption{Mean relative estimation error (left vertical axis, black) and mean standard deviation of the estimation error (right  vertical axis, red) of the 
{INT} algorithm depending on the minimum considered power (horizontal axis). }
\label{fig:err_on_min_P}
\end{figure}

However, it is worthwhile taking a closer look at the estimation errors for larger power values. To this end, we restrict the evaluation to resistors showing a certain minimum power dissipation $P_\mathrm{min}$ and plot the result in Fig.~\ref{fig:err_on_min_P}. 
In this figure, the horizontal axis shows the minimum considered resistor power $P_\mathrm{min}$, which means that for the calculation of errors \eqref{eq:absErr} and \eqref{eq:relErr}, only resistor samples with larger dissipation powers ($P_\mathrm{k,i,l}\ge P_\mathrm{min}$) are taken into account. 
Considering the black line representing the mean relative error, one can see that with rising $P_\mathrm{min}$, the estimation error drops significantly. At a minimum power of $P_\mathrm{min}\approx 300\,$mW, the mean relative error reaches approximately $10\%$. However, at the same time, the standard deviation of the estimation error (red line) further rises, {which is expected at a constant mean relative error}.

{
\subsection{Discussion}
The results show a proof of concept that in practice, accurate power estimates can be derived from IR images of electronic carrier boards. However, for future research,  some problems need to be addressed and studied. First, the proposed method relies on a manually labeled component map. 
In future work, labeling could be performed exploiting schematics or using state-of-the-art deep learning based segmentation algorithms \cite{Ren15b,He17b} tuned on electronic devices. 

Furthermore, if an accurate component map is available, the trained parameter values from the experiments performed in this paper are likely to be inaccurate for general electronic carrier boards. Although the wires and the carrier board material are probably similar in terms of thermal properties, microchips and other electrical components are expected to differ significantly from the resistors used here. Hence, for other boards, a new training of parameters should be performed. In contrast, we expect that the trained parameter values do not depend on the IR camera as long as temperature readings are sufficiently accurate. However, this is another interesting topic for future research.   

}

\section{Conclusion}
\label{sec:concl}
In this paper, we have shown that using thermal images from a consumer thermal camera, we can obtain power estimates for distinct active components on a consumer electronic circuit board. In particular, this is the first work tackling heterogeneous boards including different kinds of material with small as well as large emissivity values. To achieve power estimates, we construct a temperature-to-power conversion algorithm based on approaches from the literature, which is able to estimate the power of the distinct components. Results reveal that if the power of an active component is sufficiently large (larger than $300$\,mW), mean relative estimation errors of $10\%$ can be obtained. 

\section*{Acknowledgment}
The authors would like to thank Daniel K\"ubrich for helpful scientific discussions and Kamal Nambiar for the development of the measurement setup. 

\bibliographystyle{elsarticle-num}
\bibliography{literatureNeu}
\end{document}